# Entropy-driven excitation-energy sorting in superfluid fission dynamics

Karl-Heinz Schmidt[a] and Beatriz Jurado[b]
CENBG, CNRS/IN2P3, Chemin du Solarium B.P. 120, 33175 Gradignan, France

**Abstract:**
We study the consequences of the constant-temperature behaviour of nuclei in the superfluid regime for the exchange of excitation energy between two nuclei in thermal contact. This situation is realized at the scission configuration of fission at moderate excitation energies. It is shown that all available excitation energy is transferred to the colder fragment. This effect explains why an increase of excitation energy is translated into an increase of the number of emitted neutrons for the heavy fission fragments only. This observation remained unexplained up to now.

Most objects in nature have an approximately constant number of degrees of freedom, and their temperature, defined as the average excitation energy per degree of freedom, increases with increasing total excitation energy $E^*$ of the system. However, nuclei with moderate $E^*$ behave very differently. Experiments on nuclear level densities have shown that at least up to $E^* \approx$ 6-7 MeV the temperature of nuclei does not change with increasing $E^*$ [1]. Moreover, it was even found recently that for medium-mass nuclei the temperature stays constant up to $E^*$ = 20 MeV [2]. The main reason for this constant-temperature behaviour is that pairing correlations lead to an effective number of degrees of freedom that increases in proportion to $E^*$. Cooper pairs of neutrons and protons melt in a way that the mean energy per nucleonic excitation and thus the nuclear temperature stays constant. In nature, this behaviour appears in first-order phase transitions (e.g. solid-liquid or liquid-gas). In a mixture of two phases, like ice and water, the temperature of the mixture remains constant when energy is introduced or extracted, as long as both phases are present. Only the fractions of the two phases vary. It is of special interest to study how two quantum-mechanical objects in such a particular regime of constant temperature behave when they are in thermal contact. The scission configuration in the nuclear-fission process, where two different nuclei can exchange $E^*$ through the neck, offers a unique possibility to investigate this phenomenon.

In fission, the energy difference between the ground-state masses of the initial fissioning system and the final fission fragments, given by the $Q$ value, and the initial excitation energy of the fissioning nucleus $E^*_{CN}$, end up either in the total excitation energy (*TXE*) or in the total kinetic energy (*TKE*) of the fragments. The *TXE* is available for particle evaporation and gamma emission either before scission or from the separated fragments. In this work, we consider low-energy fission with initial excitation energies $E^*_{CN}$ up to a few MeV where evaporation and gamma emission on the fission path is considered to be weak. The same is true for neck emission of neutrons. Since fission fragments are neutron-rich, evaporation proceeds almost exclusively by neutrons. We assume that at the scission configuration the two nascent fragments have already acquired their individual properties concerning shell effects [3, 4, 5] and pairing correlations [6] and can be treated as two well defined nuclei set in thermal contact through the neck. We will now consider how the *TXE* is divided between the two nascent fragments. Following the transition-state approach of Bohr and Wheeler [7], all the available $E^*$ above the barrier height is assumed to be thermalised, that means it is, on the average, equally distributed between all available intrinsic and collective degrees of freedom. These are the single-particle excitations and the collective normal modes. The difference in potential energy between saddle and scission [8] may feed some amount of pre-scission kinetic energy in fission direction, excitations of normal collective modes and additional intrinsic excitations. We may distinguish three classes of energy, which add up to the final *TXE* of the fission fragments, according to their appearance at scission: (i) Collective excitations stored in normal modes. (ii) Intrinsic excitations by single-particle or quasi-particle excitations. (iii)

---

[a] Home institute: GSI, Planckstr. 1, D-64291 Darmstadt, Germany; e-mail: k.h.schmidt@gsi.de
[b] e-mail: jurado@cenbg.in2p3.fr

Deformation energy. The deformation energy ends up as part of the $E^*$ available when the fission fragments recover their ground-state deformations. The deformation induced in the two nascent fragments can be considered as a superposition of a macroscopic trend, caused by the mutual Coulomb repulsion of the nascent fragments, which favours a large prolate deformation around $\beta =$ 0.5 [9] and a structural influence due to shell effects. Different fission modes correspond to substantially different deformations at scission and thus to different amounts of deformation energy of the individual fragments. Theoretical arguments on the deformation of the fragments at scission can be deduced from shell-model calculations [9,10], while experimental information can be extracted from the saw-tooth-like behaviour of the neutron yields, which is thought to be mostly caused by the variation of the contribution of the deformation energy to the $E^*$ of the fragments. The division of collective excitations among the two fragments is intimately related to the nature of the specific collective mode considered. As an example, the division of $E^*$ stored in angular-momentum-bearing modes is governed by the momenta of inertia of the fragments and the conservation of total angular momentum. If the fissioning nucleus has zero angular momentum, both fragments must carry the same amount of angular momentum (in opposite direction), and, thus, the $E^*$ is inversely proportional to their moment of inertia. For these specific modes, the lighter fragment tends to carry the larger portion of $E^*$.

The division of intrinsic excitations can be derived when thermal equilibrium is assumed among the intrinsic degrees of freedom in each fragment. As said above, the nuclear level density at low $E^*$ is very well described by the constant-temperature formula:

$$\rho(E^*) \propto \exp(E^*/T) \tag{1}$$

In a recent work, Egidy et al. have obtained the following dependence of the parameter $T$ of eq. (1) with the nucleus mass number $A$ and with shell effects $S$ from a fit to available data on nuclear level densities [1]:

$$T = \frac{1}{A^{2/3}}\left(17.45 - 0.51 \cdot S + 0.051 \cdot S^2\right) \tag{2}$$

This leads to a very interesting situation for the two nascent fragments at the scission-point configuration: The level density of each fragment is represented by the constant-temperature formula (1) with a specific value of $T$ for each fragment. As a consequence, there is no solution for the division of intrinsic $E^*$ with $T_1=T_2$. As long as the fragment with the higher temperature is not completely cold, its $E^*$ is transferred to the fragment with the lower temperature. That means, a process of $E^*$ sorting takes place where all $E^*$ accumulates in the fragment with the lower value of the $T$ parameter, while the other fragment looses its entire $E^*$. According to formula (2) the heavy fragment generally has the lower $T$ and thus attracts all the $E^*$. Some deviations from the constant-temperature behaviour appear only in the range of the first quasi particle excitations [11], which might influence the energy-sorting mechanism in its final phase.

Due to the influence of shell corrections on $T$, see eq. (2), the direction of the energy transfer may be reversed if the heavy fragment is stabilized by a strong shell effect. This may be expected in the standard I (SI) fission channel, which is characterised by the formation of a heavy fragment close to the doubly magic $^{132}$Sn. The flow of $E^*$ from the hot fragment to the cold fragment can be seen as a way for the entire system made of the two nascent fragments in contact to maximize the number of occupied states or its entropy. The number of available states of the light nucleus or closed-shell nucleus is small compared to that of the complementary fragment. Therefore, the situation in which the light nucleus or the closed-shell nucleus has part of the $E^*$ leads to a smaller entropy than the situation in which the entire $E^*$ is transferred to the heavy or the non-closed-shell nucleus which has considerable more available states.

The number of evaporated neutrons as a function of the fragment mass is directly related to the $E^*$ of the fragment and therefore should clearly reflect the peculiar situation of the full transfer of the intrinsic $E^*$ to the cold fragment. The neutron-induced fission of $^{237}$Np has been studied very carefully at two different neutron energies [12]. Fig. 1 shows the average number of evaporated neutrons as a function of the fragment mass. As mentioned above, the well known saw-tooth-like behaviour of this curve is attributed to the deformation energy. The minimum close to $A$=130 is due to the shell closures $N$=82, $Z$=50 that lead to spherical fission fragments. An increase of incident neutron energy translates into an increase of $E^*_{CN}$. The increase of the emitted neutrons near symmetry for 110<$A$<130 with incident neutron energy is caused by the increase of the yield of the super long (SL) mode which is related to well deformed fission fragments. For more asymmetric mass splits outside this range, we observe a very peculiar feature: Interestingly, Fig. 1 shows that the increase of $E^*$ leads to an increase of the number of evaporated neutrons for the heavy fragment, only. Actually, a quantitative analysis of the data reveals that all of the increased $E^*$ appears in the heavy fragment. This observation is rather general as it was also found for other fissioning systems such as $^{233}$U and $^{238}$U and other incident particles like protons [13, 14, 15, 16]. However, no clear explanation has yet been found for this effect. The reason is that all the work [17, 18, 19] done to study the partition of intrinsic excitation energy between fission fragments is based on a wrong description of the level density of the fragments. In fact, up to now one has assumed that the level density at low $E^*$ is well described by the analytical formula of Bethe [20]

$$\rho(E^*) \propto \exp\left(2\sqrt{aE^*}\right) \qquad (3)$$

where $a$ is the level-density parameter which is proportional to the mass number $A$ of the nucleus. The latter formula is based on an equidistant single-particle level scheme. The temperature $T$ is given by the inverse logarithmic slope of $\rho(E^*)$:

$$\frac{1}{T} = \frac{d\ln(\rho(E^*))}{dE^*} \qquad (4)$$

which leads to the well-known relation

$$E^* = aT^2 \qquad (5)$$

between $E^*$ and the temperature $T$ of the system. If the temperatures of the two nascent fragments are required to be equal, we obtain an intrinsic $E^*$ division in proportion of the mass ratio of the fragments

$$\frac{E_1^*}{E_2^*} = \frac{A_1}{A_2} \qquad (6)$$

The tendency to realize relation (6) has been confirmed empirically in many binary reactions involving relatively high $E^*$ [21], although also deviations from full equilibration were observed due to insufficient reaction time [22]. However, this expression is not applicable at low $E^*$ and fails to explain the observation presented in Fig. 1 that all the increase in $E^*$ is found in the heavy fragment. Actually, this effect is a direct consequence of the different constant temperatures of the two fragments at scission. According to eq. (2), the temperature of the heavy fragment, in the absence of strong shell effects, is always lower than the temperature of the light fragment. Therefore, the heavy fragment will absorb the entire available intrinsic $E^*$ and evaporate more neutrons. When the heavy fragment is close to the $^{132}$Sn shell closures, its temperature is increased. According to eq. (2) the value amounts to $T_{heavy}$=0.83 MeV (with $A_{heavy}$=130 and $S_{heavy}$ = -5MeV

[23]). If the shell effect of the complementary light fragment is zero, its temperature amounts to only $T_{light}$=0.77 MeV (with $A_{light}$=108 and $S_{light}$= 0), and, thus, the direction of the energy-sorting mechanism will be reversed. In this case, the light fragment receives the entire intrinsic $E^*$. The dip around $A$=130 and the peak related to the complementary fragment around $A$=108 may be an indication for this feature. Note that the magnitude with which the effect of the shell closure is reflected by the data depends also on the relative yield at $A$=130 of the S1 fission channel with respect to the SL and the Standard 2 (S2) channels. The deformed shell closure around $N$=88, assumed to be responsible for the S2 fission channel, is weaker [9, 10] than the $^{132}$Sn shell closure and related to more asymmetric splits than the S1 channel. If we tentatively assume $S_{heavy}$ = -1 MeV (-4 MeV), this leads to $T_{heavy}$ = 0.68 MeV (0.75MeV) and $T_{light}$= 0.82 MeV where we have taken $A_{heavy}$= 140, $A_{light}$= 98 and $S_{light}$= 0. Therefore, in this case the shell closure is not strong enough to reverse the direction of the flow of $E^*$. As for the symmetric fission channel SL, for S2 the total $E^*$ is found in the heavy fragment.

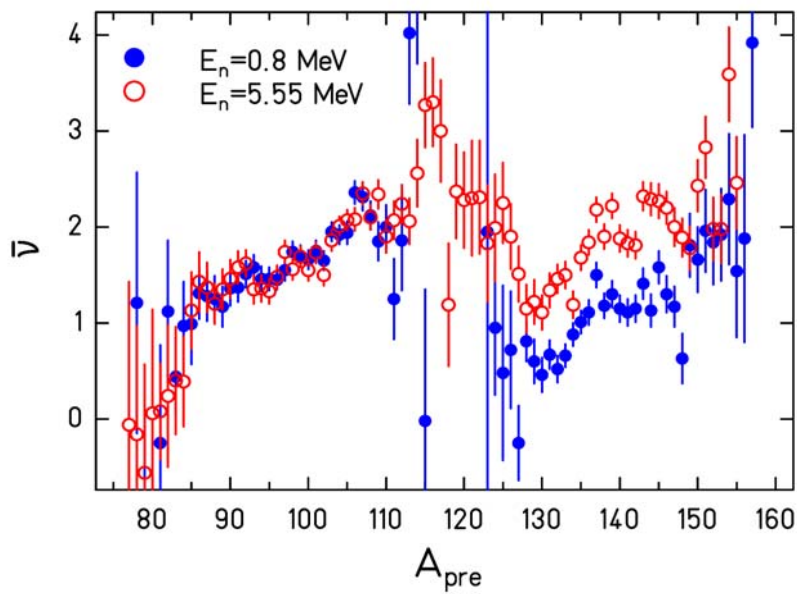

**Figure 1 :** (Colour online) Average number of prompt neutrons as a function of the primary fragment mass for the neutron-induced fission of $^{237}$Np at two incident neutron energies, data taken from ref. [12].

We would like to stress that our argumentation is based on the same assumptions as other work that investigates the sharing of intrinsic $E^*$ at scission [17, 18, 19]. That is, we have assumed independent fission fragments and thermal equilibration between the fragments at scission. What is substantially different in our approach is that we use the constant-temperature level density which correctly describes the behaviour of nuclei at moderate $E^*$ and not the commonly used Fermi gas level density of eq. (3) derived by Bethe [20] which is only valid at high $E^*$.

In conclusion, at low excitation energy $E^*$ nuclei are peculiar systems, characterised by a phase transition from superfluidity to normal-liquid behaviour. As is typical for first-order phase transitions, their temperature remains constant with increasing $E^*$. In this sense, the nuclear superfluid to normal-liquid phase transition seems to behave like a first-order phase transition. The very special feature of this phenomenon in nuclei is that the constant-temperature regime reaches down to zero energy. The scission configuration of the fission process offers the unique possibility to investigate how two different nuclei in this special regime of constant temperature share the available intrinsic excitation when they are in thermal contact. We have shown for the first time that in this regime we reach a peculiar state of thermal equilibrium at scission in which the temperatures of the nascent fragments remain different in spite of the flow of $E^*$ from the hot to the cold fragment. Rather unexpectedly, this implies that the total amount of intrinsic $E^*$ available at scission is found in the fragment with the lower temperature. This entropy-driven $E^*$-sorting process appears to have similarities with Maxwell's demon [24] on the nucleonic level. However, the phenomenon is fully compatible with the second law of thermodynamics. This $E^*$-sorting effect explains very easily an issue that remained unsolved up to present when comparing the number of emitted neutrons as a function of fragment mass for different initial excitation energies. It was observed in asymmetric mass splits that the increase of intrinsic $E^*$ of the fissioning nucleus appears as an increase of $E^*$ in the heavy fission fragments, only. Indeed, the temperature of the heavy fission fragments is generally lower than that of the light ones. Therefore, all the intrinsic $E^*$ is cumulated in the heavy fragment. Our work shows that the behaviour of highly excited systems according to the Fermi-gas level density, where the total $E^*$ is shared by the fragments in proportion of their masses, is strongly violated at moderate $E^*$. Our finding has important consequences for the understanding of fission. Many conclusions that result from a wrong assumption on how the total intrinsic excitation energy is shared between the fission fragments should be revisited.

**Acknowledgments**
This work was supported by the EURATOM 6. Framework Programme "European Facilities for Nuclear Data Measurements" (EFNUDAT), contract number FP6-036434. We thank F. Käppeler for providing us with the numerical values of the neutron-yield data.

1. T. von Egidy, D. Bucurescu, Phys. Rev. C **72**, 044311 (2005).
2. A. V. Voinov et al., Phys. Rev. C **79**, 031301(R) (2009).
3. U. Mosel, H. Schmitt, Phys. Rev. C **4**, 2185 (1971).
4. U. Mosel, H. Schmitt, Nucl. Phys. A **165**, 73 (1971).
5. J. Maruhn, W. Greiner, Z. Phys. **251**, 211 (1972).
6. H. J. Krappe, S. Fadeev, Nucl. Phys. A **690**, 431 (2002).
7. N. Bohr, J. A. Wheeler, Phys. Rev. **56**, 426 (1939).
8. M. Asghar, R. W. Hasse, J. Phys. Colloques **45**, C6-455 (1984).
9. B. D. Wilkins, E. P. Steinberg, R. R. Chasman, Phys. Rev. C **14**, 1832 (1976).
10. I. Ragnarsson, R. K. Sheline, Phys. Scr. **29**, 385 (1984).
11. U. Agvaanluvsan et al., Phys. Rev. C **79**, 014320 (2009).
12. A. A. Naqvi, F. Käppeler, F. Dickmann, R. Müller, Phys. Rev. C **34**, 218 (1986).
13. S. C. Burnett, R. L. Ferguson, F. Plasil, H. W. Schmitt, Phys. Rev. C **3**, 2034 (1970).
14. C. J. Bishop, R. Vandenbosch, R. Aley, R. W. Shaw Jr., I. Halpern, Nucl. Phys. A **150**, 129 (1970).
15. R. Müller, A. A. Naqvi, F. Käppeler, F. Dickmann, Phys. Rev. C **29**, 885 (1984).
16. M. Strecker, R. Wien, P. Plischke, W. Scobel, Phys. Rev. C **41**, 2172 (1990).
17. D. G. Madland, J. R. Nix, Nucl. Sci. Eng. 81, 213 (1982).
18. S. Lemaire, P. Talou, T. Kawano, M. B. Chadwick, D. G. Madland, Phys. Rev. C **72**, 024601 (2005).
19. N. V. Kornilov, F.-J. Hambsch, A. S. Vorobyev, Nucl. Phys. A **789**, 55 (2007).
20. H. A. Bethe, Phys. Rev. **50**, 332 (1939).
21. J. Toke, W. U. Schröder, Annu. Rev. Nucl. Sci. **42**, 401 (1992).
22. S. Piantelli et al., Phys. Rev. C **78**, 064605 (2008).
23. V. V. Pashkevich, Nucl. Phys. A **169**, 275 (1971).
24. J. Earman, J. D. Norton, Stud. Hist. Phil. Mod. Phys. **29**, 435 (1998).